\documentclass[preprints,article,accept,moreauthors,pdftex]{Definitions/mdpi} 

\usepackage{bm}
\usepackage{bbm}
\usepackage{listings}
\usepackage{subfig}
\usepackage{amsmath}
\usepackage{amssymb}
\usepackage{booktabs}
\usepackage{hyperref}
\usepackage{fourier}
\usepackage[T1]{fontenc}
\usepackage[]{avant}
\usepackage{longtable}
\usepackage[ruled, lined, linesnumbered, commentsnumbered, longend]{algorithm2e}

\hypersetup{colorlinks,citecolor=blue}

\firstpage{1} 
\pubvolume{xx}
\issuenum{1}
\articlenumber{1}
\pubyear{2023}
\copyrightyear{2023}
\history{}

\newcommand*\diff{\mathop{}\!\kern0pt\mathrm{d}}

\Title{Instabilities of Super-Time-Stepping Methods on the Heston Stochastic Volatility Model}
\Author{Fabien Le Floc'h}

\AuthorNames{Fabien Le Floc'h}

\address{fabien@2ipi.com}
\abstract{This note explores in more details instabilities of explicit super-time-stepping schemes, such as the Runge-Kutta-Chebyshev or Runge-Kutta-Legendre schemes, noticed in the litterature, when applied to the Heston stochastic volatility model. The stability remarks are relevant beyond the scope of super-time-stepping schemes.}
\keyword{Finite difference method; stability; quantitative finance; stochastic volatility}

\begin{document}
	
	\section{Introduction}
	Explicit super-time-stepping schemes offer an interesting alternative to classic implicit discretization schemes for advection-diffusion partial differential equations (PDEs), especially on multi-dimensional problems, such as the pricing of financial derivatives under the Heston stochastic volatility model \citep{heston1993closed}.
	A-stable or L-stable implicit schemes may indeed be slow, because a linear system must be solved at each time-step. Hence the popularity of various splitting schemes, and especially the alternative direction implicit (ADI) variety, to make the multiple resulting
	 linear systems faster to solve, at the cost of a greater complexity. On the other side, the explicit Euler scheme, with its limited stability region, is well known to require too many time-steps to be practical for advection-diffusion PDEs. The super-time-stepping schemes allow to circumvent this limitation.
	
	\citet{foulon2010adi} noticed instabilities of the Runge-Kutta-Chebyshev (RKC) super-time-stepping scheme with relatively large shift ($\epsilon=10$) on the Heston PDE in the context of a low vol-of-variance parameter ($\sigma=4\%$). This is somewhat surprising, since the large shift increases the stability and damping properties of the scheme significantly \citep{verwer2004rkc}. \citet{lefloch2019numerical} did not notice such instabilities, on the same problem, with the shifted RKC scheme, or with the Runge-Kutta-Legendre (RKL) scheme. 	\citet{o2013pricing} also applied a RKC scheme to the Heston PDE with a low vol-of-variance $\sigma=1\%$, without issues.

	We aim here at clarifying the discrepancy and we look in more details at the stability of the RKC, RKL or Runge-Kutta-Gegenbauer (RKG) schemes on this problem.
	
	\section{The Heston PDE and its discretization}
	\subsection{The Heston PDE}
	In the stochastic volatility model of \citet{heston1993closed}, the asset $X$ follows
	\begin{subequations}
		\begin{align}
			\diff X(t) &= (r(t)-q(t)) X(t)dt+\sqrt{V(t)} X(t) \diff W_X(t)\,,\label{eqn:heston_X}\\ 
			\diff V(t) &= \kappa \left(\theta - V(t)\right) + \sigma \sqrt{V(t)} \diff W_V(t)\,,\label{eqn:heston_V}
		\end{align}
	\end{subequations}
	with $W_X$ and $W_V$ being two Brownian motions with correlation $\rho$, and $r, q$ the instantaneous growth and dividend rates.
	
	The corresponding PDE for the option price $f$ reads
	\begin{equation}
		\frac{\partial f}{\partial t} = \frac{v x^2}{2}\frac{\partial^2 f}{\partial x^2} + \rho\sigma x v \frac{\partial^2 f}{\partial x \partial v} + \frac{\sigma^2 v}{2} \frac{\partial^2 f}{\partial v^2} + (r-q)x\frac{\partial f}{\partial x} + \kappa(\theta - v)\frac{\partial f}{\partial v} - r f\,,\label{eqn:heston_pde}
	\end{equation}
	for $0 \leq t \leq T$, $x > 0$, $v > 0$, with initial condition  $f(T, x, v) = F(x)$.
	
	\subsection{Discretization}
	We follow \citet{lefloch2019numerical,foulon2010adi} and solve the  PDE on a truncated domain $[x_{\min}, x_{\max}]\times[v_{\min},v_{\max}]$. In \citep{lefloch2019numerical} the bounds read
	\begin{align*}
		x_{\min} = 0\,,&\quad x_{\max} =K e^{+4\sqrt{\theta T}}\,.
	\end{align*}
	where $K$ is the option strike price, while in \cite{foulon2010adi}, we have $x_{\max} = 8K$.
	
	For the $v$ domain, let $\Phi_{\chi}(y,d_{\chi},\lambda_{\chi})$ be the cumulative distribution for the non-central chi-square distribution with $d_{\chi}$ degrees of freedom and non-centrality parameter $\lambda_{\chi}$. The distribution of the variance process $V(T)$ conditional on $V(0)$ is known \citep{cox1985theory}, and \citet{lefloch2019numerical}  choose
	\begin{align*}
		v_{\min} = 0\,,\quad v_{\max} = \Phi_{\chi}^{-1}(1-\epsilon_v, d_{\chi}, v_0 n_\chi) \frac{e^{-\kappa T}}{n_{\chi}}\,,
	\end{align*}
	with $d_{\chi} = 4\frac{\kappa\theta}{\sigma^2}$, $n_\chi = 4\kappa \frac{e^{-\kappa T}}{\sigma^2\left(1-e^{-\kappa T}\right)}$, and $\epsilon_v=10^{-4}$, while \cite{foulon2010adi} choose $v_{\max}=5$.

	At $v =v_{\max}$, \citet{lefloch2019numerical} follow \citet[p. 385-386]{andersen2010interestcv} and let the price be linear in the variance dimension:
	\begin{equation}\frac{\partial^2 f}{\partial v^2}(x,v,t) = \frac{\partial^2 f}{\partial x\partial v}(x,v,t) = 0\,,\label{eqn:vmin_boundary}\end{equation}
	for $x \in (x_{\min},x_{\max})$. 
 When $v_{\min} = 0$, the exact boundary condition at $v=v_{\min}=0$ corresponds to the PDE obtained by setting $v=0$.
	
	At $x_{\max}$ and $x_{\min}$, we consider that the value is linear along $x$, which leads to
	\begin{align}
		\frac{\partial^2 f}{\partial x^2}(x,v,t) =\frac{\partial f}{\partial v}(x,v,t) = 0\,, \label{eqn:linear_bc_sv}
	\end{align}
	for  $v \in [v_{\min},v_{\max}]$  and $x \in \left\{x_{\min},x_{\max}\right\}$.
	
	With a second-order central discretization of the derivatives,  the explicit step involved at each stage of the super-time-stepping scheme reads
	\begin{align*}
		\hat{f}^\eta_{i,j} =& \hat{g}^{\eta-1}_{i,j} +\bar{\lambda}_\eta a_{i,j}\hat{f}^{\eta-1}_{i-1,j} + \bar{\lambda}_\eta b_{i,j} \hat{f}^{\eta-1}_{i,j} + \bar{\lambda}_\eta c_{i,j}\hat{f}^{\eta-1}_{i+1,j} + \bar{\lambda}_\eta d_{i,j}\hat{f}^{\eta-1}_{i,j-1} + \bar{\lambda}_\eta e_{i,j}\hat{f}^{\eta-1}_{i,j+1}\\
		&+\bar{\lambda}_\eta \omega_{i,j} \left(\hat{f}^{\eta-1}_{i+1,j+1} - \hat{f}^{\eta-1}_{i+1,j-1} - \hat{f}^{\eta-1}_{i-1,j+1} + \hat{f}^{\eta-1}_{i-1,j-1}\right)\,,
	\end{align*}
	with
	\begin{subequations}
		\begin{align}
			a_{i,j} =& -\frac{k}{h_i\left(h_{i+1}+h_i\right)}\left(\mu_{i} h_{i+1} x_i - \beta^x_{i,j} v_j x_i^2\right) \label{eqn:stencil_eq1}\,,\\
			b_{i,j} =& - k \left(r_{i}+ \frac{\mu_{i}(h_i-h_{i+1})+\beta^x_{i,j} v_j x_i^2 }{h_i h_{i+1}} +  \frac{\kappa(\theta-v_j)(w_j-w_{j+1})+\beta^v_{i,j} v_j \sigma^2 }{w_j w_{j+1}}\right)\,, \label{eqn:stencil_eq2}\\
			c_{i,j} =& \frac{k }{h_{i+1}\left(h_{i+1}+h_i\right)}\left(\mu_{i}h_i x_i+ \beta^x_{i,j} v_j x_i^2\right) \label{eqn:stencil_eq3}\,,\\
			d_{i,j} =& -\frac{k  }{w_{j}\left(w_{j+1}+w_j\right)}\left(\kappa (\theta-v_j)w_{j+1} - \beta^v_{i,j} v_j \sigma^2\right) \label{eqn:stencil_eq4}\,,	\\
			e_{i,j} =& \frac{k  }{w_{j+1}\left(w_{j+1}+w_j\right)}\left(\kappa (\theta-v_j)w_j + \beta^v_{i,j} v_j \sigma^2\right) \label{eqn:stencil_eq5}\,,\\
			\omega_{i,j} =& \frac{k \rho\sigma  x_i v_j}{(h_i+h_{i+1})(w_j + w_{j+1})} \label{eqn:stencil_eq6}\,,
		\end{align}
	\end{subequations}
	and $h_i = x_i - x_{i-1}$, $w_j = v_j-v_{j+1}$, for $i=1,...,m-1$, and $j=1,...,n-1$. The classic central discretization correspond to the choice $\beta^x_{i,j}=\beta^v_{i,j}=1$.
	In order to simplify the notation, we dropped the time index and all the coefficients $\mu_i, r_i, k$ involved are understood to be taken at a specific time-step. The index $j$ applies to the variance dimension instead of the time dimension in the previous sections of this paper.
	
	The cell P\'eclet number $P$ is the ratio of the advection coefficient towards the diffusion coefficient in a cell \citep{hundsdorfer2013numerical}. When the P\'eclet condition $P \leq 2$ does not hold, the stability of the finite difference scheme is not guaranteed anymore: the solution may explode.  Here, the cell P\'eclet number for each dimension is 
	\begin{align}
		P^{x}_{i, j}(\beta^x_{i,j}) = \frac{2h_i}{\beta^x_{i,j}  v_j x_i} \left(r_i-q_i\right)&\,,\quad P^{v}_{i,j}(\beta^v_{i,j}) =\frac{2 w_j\kappa(\theta-v_j)}{\beta^v_{i,j} \sigma^2 v_j}\,.
	\end{align}                                   
	The P\'eclet conditions $P^{x}_{i,j} < 2$ and $P^{v}_{j} < 2$ do not necessary hold with typical values for the Heston parameters. This happens when $v_j$ is very small, which is generally the case for the first few indices $j$.  In order to ensure that the P\'eclet conditions hold, \citet{lefloch2019numerical} use the exponential fitting technique of \citet{allen1955relaxation, il1969differencing} when $P^{x}_{i,j} \geq 2$ as well as when $P^{v}_{i,j} \geq 2$. It consists in using the coefficients \begin{align}
		\beta^{x}_{i,j} = \frac{P^{x}_{i,j}(1)}{2\tanh\left(\frac{P^{x}_{i,j}(1)}{2}\right)} &\,,\quad \beta^{v}_{i,j} = \frac{P^{v}_{i,j}(1)}{2\tanh\left(\frac{P^{v}_{i,j}(1)}{2}\right)}\,,\label{eqn:peclet}
	\end{align}
	instead of $\beta^{x}_{i,j}=\beta^{v}_{i,j}=1$. 
	
	The boundary conditions, discretized with order-1 forward and backward differences, read
	\begin{align*}
		a_{i,0} &= -\frac{k}{h_i\left(h_{i+1}+h_i\right)}\left(\mu_{i} h_{i+1} x_i - \beta^x_{i,0} v_0 x_i^2\right)\,,\\
		c_{i,0} &= \frac{k  }{h_{i+1}\left(h_{i+1}+h_i\right)}\left(\mu_{i}h_i x_i+ \beta^x_{i,0} v_j x_i^2\right) \,,\\
		b_{i,0} &= - k \left(r_{i}+ \frac{\mu_{i}(h_i-h_{i+1})+\beta^x_{i,0} v_0 x_i^2 }{h_i h_{i+1}} +  \frac{\kappa(\theta-v_0)}{w_1}\right)\,,\\
		d_{i,0} &= 0\,,	\quad e_{i,0} = \frac{k \kappa (\theta-v_0) }{w_{1}}\,,\quad w_{i,0} = 0\,,\\
		a_{i,n} &= -\frac{k}{h_i\left(h_{i+1}+h_i\right)}\left(\mu_{i} h_{i+1} x_i - \beta^x_{i,n} v_n x_i^2\right)\,,\\
		c_{i,n} &= \frac{k  }{h_{i+1}\left(h_{i+1}+h_i\right)}\left(\mu_{i}h_i x_i+ \beta^x_{i,n} v_j x_i^2\right) \,,\\
		b_{i,n} &= - k \left(r_{i}+ \frac{\mu_{i}(h_i-h_{i+1})+\beta^x_{i,n} v_n x_i^2 }{h_i h_{i+1}} -  \frac{\kappa(\theta-v_n)}{w_{n-1}}\right)\,,\\
		d_{i,n} &= -\frac{k \kappa (\theta-v_n) }{w_{n-1}}\,,	\quad e_{i,n} = 0\,,\quad \omega_{i,n} = 0\,,
	\end{align*}
	for $i=1,...,m-1$, and
	\begin{eqnarray*}
		a_{0,j}=0\,, \quad {b}_{0,j} = - k\left(r_{0}+\frac{\mu_{0}x_0}{h_1}\right)\,,\quad  {c}_{0,j} = k \frac{\mu_{0}x_0}{h_1}\,,\quad d_{0,j}=e_{0,j}=\omega_{0,j}=0\,,\\
		{a}_{m,j}  = - k\frac{\mu_{m}x_m}{h_{m}}\,,\quad  {b}_{m,j} = - k\left(r_{m}-\frac{\mu_{m}x_m}{h_{m}}\right)\,,\quad c_{m,j} = d_{m,j}=e_{m,j}=\omega_{m,j}=0\,,
	\end{eqnarray*}
	for $j=0,...,n$.

	\citet{o2013pricing}  rely on one-sided upwinding anywhere the PDE becomes convection dominated.
	In contrast, \citet{foulon2010adi} use three-point upwinding finite difference approximations at $v=v_{\min}$ and for $v > 1$. At $x_{\min}$ and $x_{\max}$ slightly different boundary conditions are used, but those do not matter for the stability of the studied example.
	
	\subsection{Grid}
	\citet{foulon2010adi} use a non-uniform grid, with points concentrated around $x=K$ and $v=v_{\min}$ through a the hyperbolic transformation presented in \citep{tavella-pricing}. For the $x$ coordinate, the transformation reads
\begin{align}
	x( \eta ) &= K + \lambda_x \sinh((c_{x,2}-c_{x,1})\eta+c_{x,1})\,,\label{eqn:S_transform}\end{align}
with $c_{x,1} = \sinh^{-1}\left(\frac{x_{\min}-K}{\lambda_x}\right)$, $c_{x,2} = \sinh^{-1}\left(\frac{x_{\max}-K}{\lambda_x}\right)$, $\eta$ uniform in $[0,1]$ and $\lambda_x = K/5$. The same transformation is used for $v$ (replacing $x$ by $v$ and $K$ by $v_{\min}$) with $\lambda_v = v_{\max}/500$.
	
	\citet{lefloch2019numerical} concentrates the point around $v=v_0$ instead with a milder streching $\lambda_v=2 v_0$.
\citet{o2013pricing} rely on different stretchings to ensure that the discretized matrix is still an M-matrix (for the $x$ coordinate towards zero), still concentrate points around $x=K$, and double the amount of points close to $v_{\min}$ compared to larger variances.

	In our numerical examples for Heston, we will follow \citet{foulon2010adi} for the grid boundaries and stretching.
	
	\section{Root of the discrepancy}
\begin{figure}[h]
	\centering{
		\includegraphics[width=.7\textwidth]{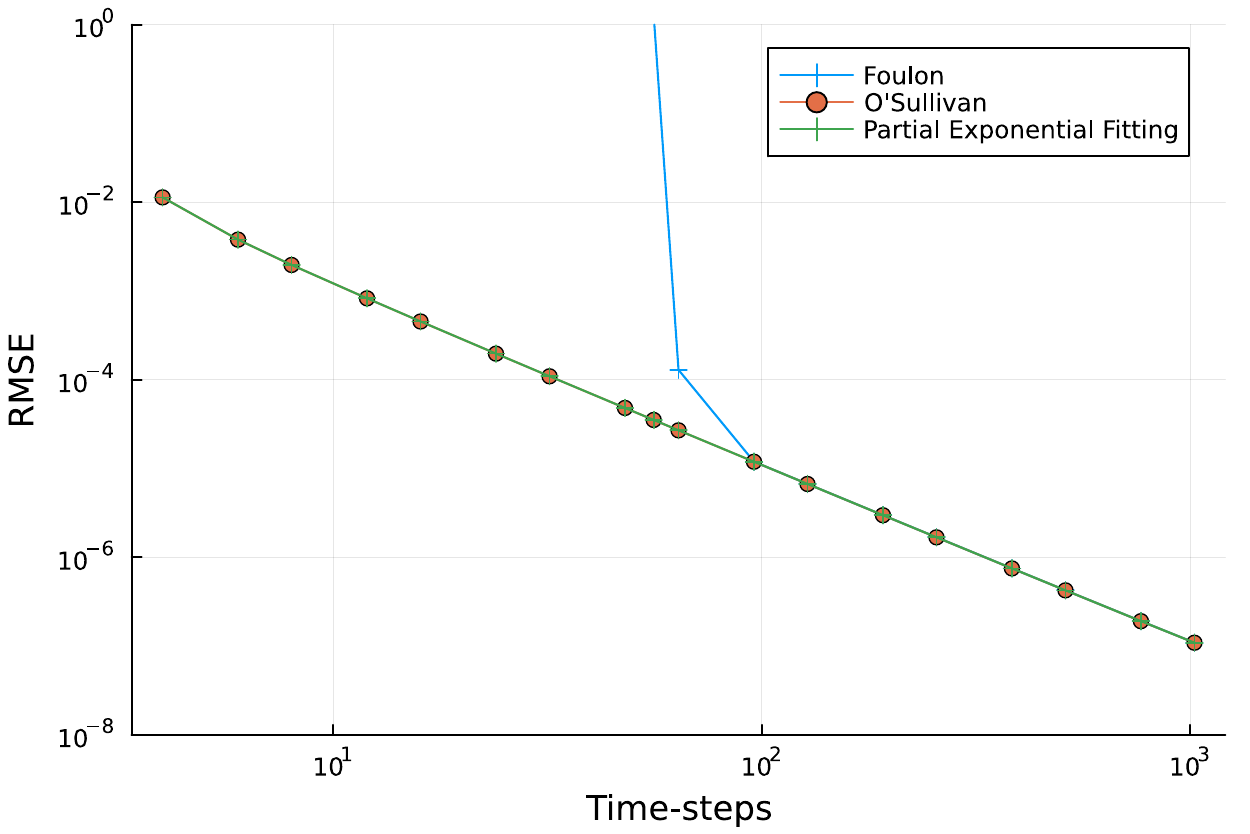}}
		\caption{Convergence in time of the RKC scheme with $\epsilon=10$, with the different choices of upwinding with $m=100$, $n=50$. The reference prices are the obtained with the Craig-Sneyd scheme on 16000 time-steps, using the same discretization in space and the same upwinding choice.\label{fig:rkg_time_convergence.pdf}}
\end{figure}

The apparent discrepancy lies in the finite difference discretization choices, and particularly the upwinding. In \citep{foulon2010adi}, three points upwinding is used at $v=0$ and for $v >1$ while in \citep{lefloch2019numerical}, exponential fitting is used when the Peclet number $P > 2$ and single-sided differences are used at the boundaries $x_{\min}, x_{\max}, v_{\min}, v_{\max}$. If we restrict exponential fitting to the same region as used in  \citep{foulon2010adi}, we end up with similar instabilities for the RKC and RKL schemes as noticed in the latter paper with the Heston parameters given in Table \ref{tbl:foulon2}. 
\begin{table}[h]
	\caption{Parameters for the case II of \citet{foulon2010adi}.\label{tbl:foulon2}}
	\centering{
	\begin{tabular}{cccccccccc}\toprule
		\multicolumn{5}{c}{Heston} & \multicolumn{3}{c}{Market} & \multicolumn{2}{c}{Option}\\\cmidrule(lr){1-5}\cmidrule(lr){6-8}\cmidrule(lr){9-10}
		$V(0)$ & $\theta$ & $\kappa$ & $\sigma$ & $\rho$ & $r$ & $q$ & $X(0)$ & $K$ & $T$ \\\midrule
		0.12 & 0.12 & 3.0 & 0.04 & 0.6 & 0.01 & 0.04 & 100 & 100 & 1\\\bottomrule
	\end{tabular}}
\end{table}
We label this upwinding as "Foulon" in Figure \ref{fig:rkg_time_convergence.pdf},
 even though it is slightly different from the paper (exponential fitting vs. three points upwinding but applied on the same region). There is a clear explosion of the error in time\footnote{The error in time is the root mean square error between the scheme values using $l$ time-steps at a subset of the grid points and the values obtained by a reference scheme on the same grid along $(x,v)$ but using many more time-steps. See \cite{foulon2010adi} for a precise definition.} when the number of time-steps is below 100. In contrast, with the partial exponential fitting (applied anywhere the P\'eclet number is larger than two) or the one-sided upwinding from \citet{o2013pricing} (applied anywhere the PDE becomes convection dominated), the RKC scheme works well regardless of the number of time-steps.

\begin{figure}[h]
	\centering{
		\includegraphics[width=.7\textwidth]{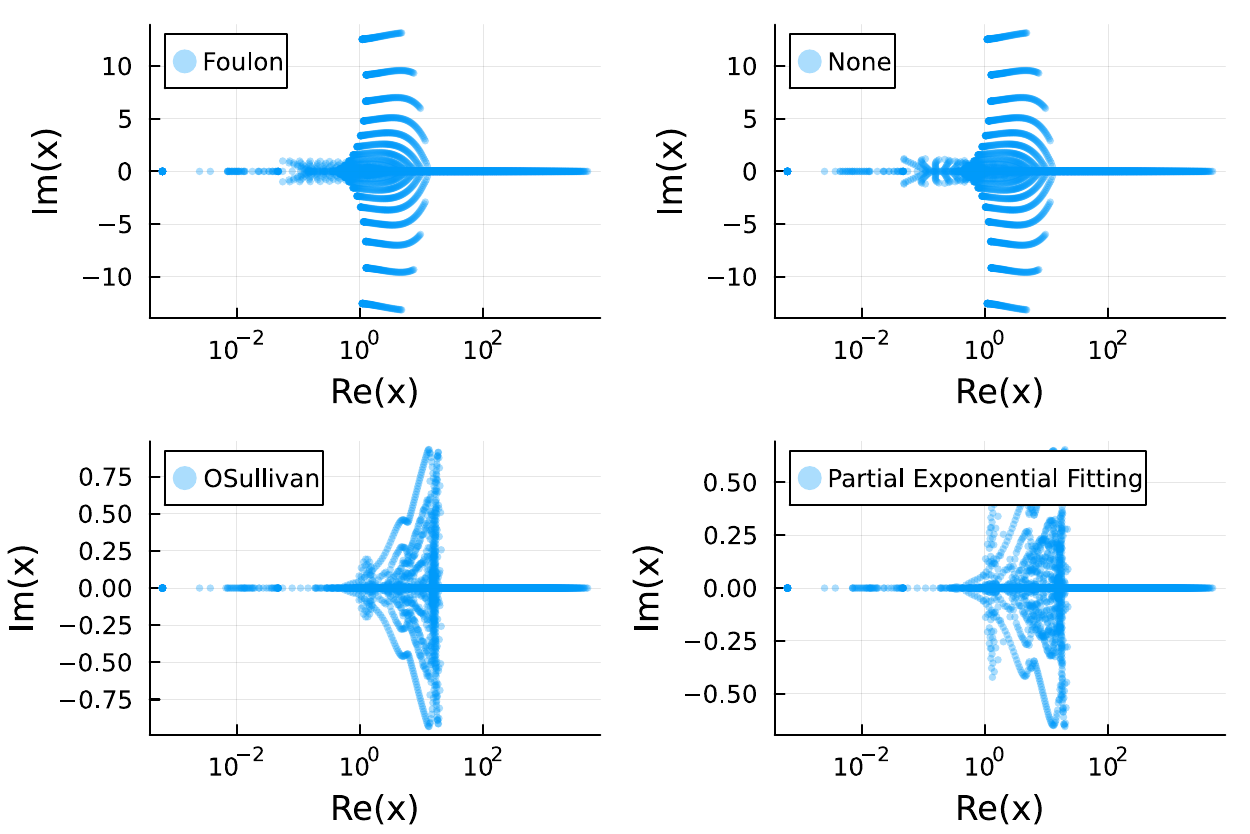}}
	\caption{Eigenvalues of the discretization matrix with $l=16, m=100, n=50$ using different upwinding choices. Note the imaginary axis range difference.\label{fig:eigenvalues_N16M100L50.pdf}}
\end{figure}

With the Foulon upwinding, the eigenvalues have indeed  larger imaginary parts. In fact, the upwinding there is not having a noticeable impact on the eigenvalues and stability  (Figure \ref{fig:eigenvalues_N16M100L50.pdf}). 
The partial exponential fitting and O'sullivan approaches decrease the maximum imaginary part by a factor larger than 10.

The instabilities should not be too surprising since the advection may dominate for the coordinate $v$ and $v$ close to zero.                     
In case of an advection equation, it is well known that the explicit Euler time-stepping with central differencing is unstable \citep{hundsdorfer2013numerical}. On the example considered, the explicit Euler scheme is actually convergent, likely because the oscillations are restricted to a specific zone and do not propagate to the regions where the diffusion is larger. For the RKC scheme, \citet{verwer2004rkc} advise the use of upwinding for advection.
On the example considered the RKC scheme with no upwinding may become stable again if we further increase the damping shift to $\epsilon= 1000$ at the cost of twice the number of stages, or if we increase the number of time-steps. Clearly then, the use of upwinding in the full region where $P>2$ is more appropriate. 

\begin{figure}[h!]
	\centering{
		\subfloat[RKC with damping shift $\epsilon=10$. No oscillations are visible near $v=0$.]{
			\includegraphics[width=.8\textwidth]{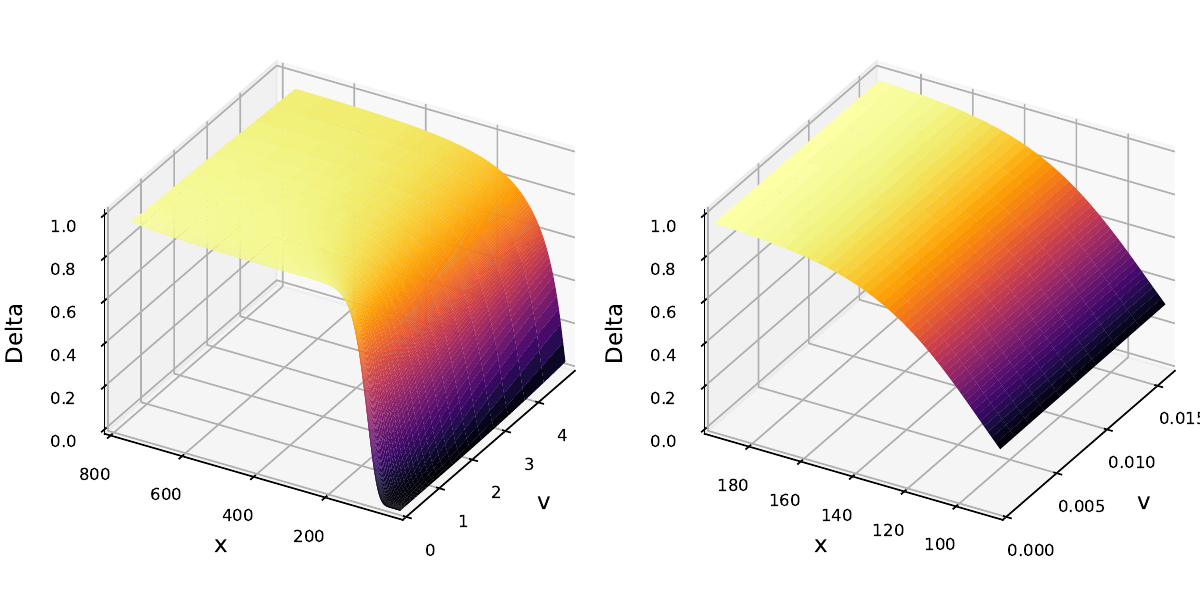}}\\
		\subfloat[RKL. Large oscillations are visible near $v=0$.]{
			\includegraphics[width=.8\textwidth]{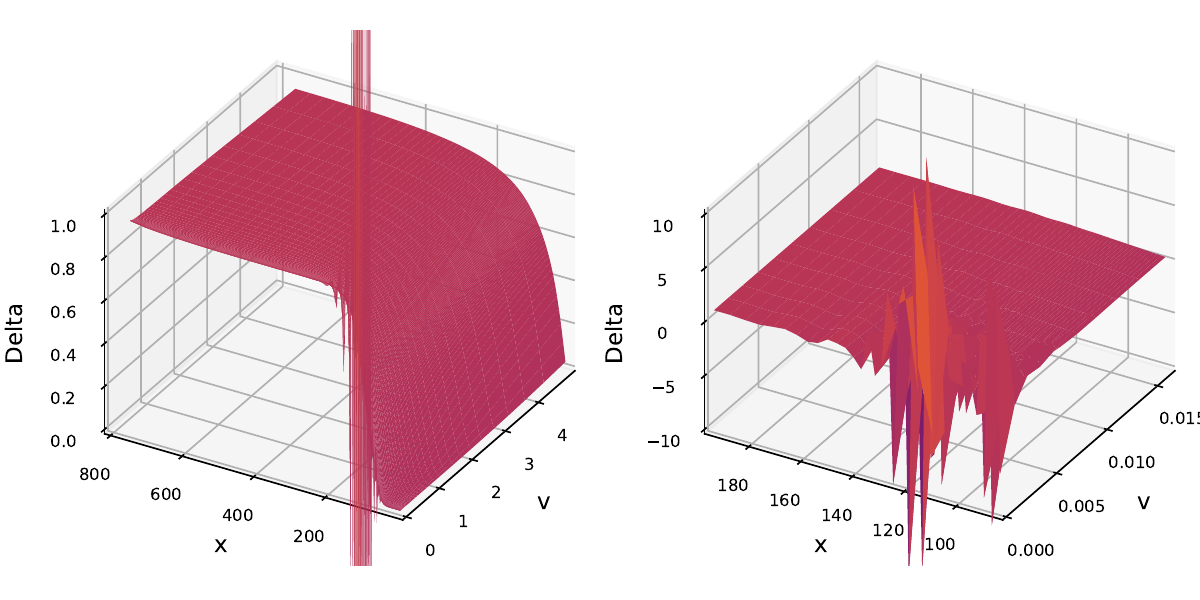}}\\
		\subfloat[RKG. Very small oscillations are visible at $v=0$.]{
			\includegraphics[width=.8\textwidth]{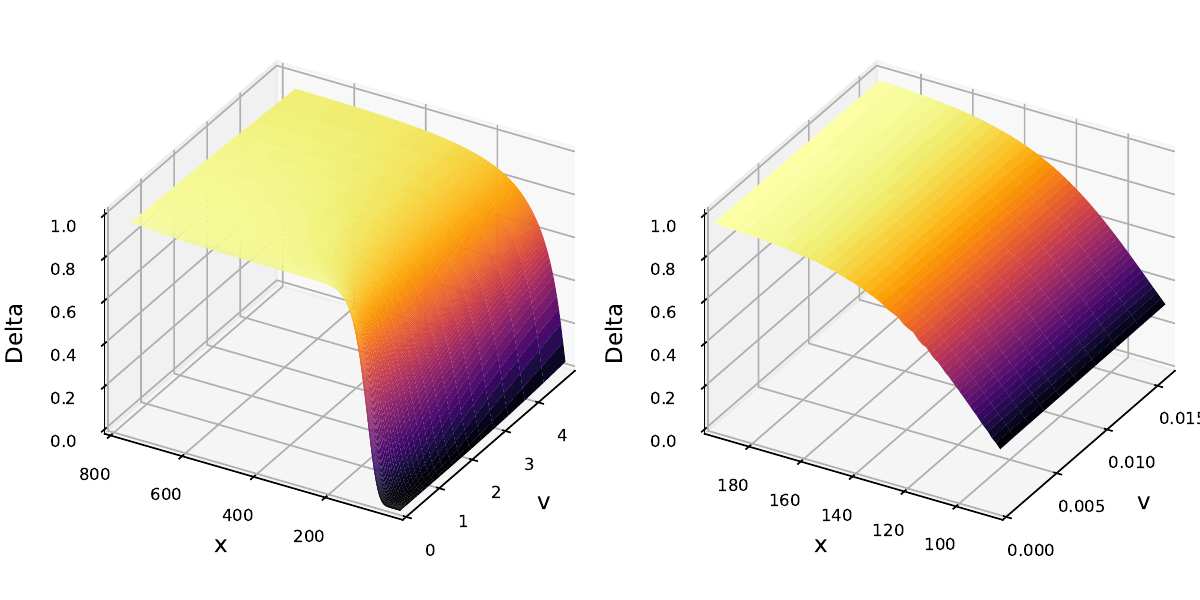}}		
		\caption{Delta by forward difference on different super-time-stepping schemes for $l=10$ time-steps  and partial exponential fitting on the grid $m=100,n=50$ (Left: full grid, Right: zoom on small $v$).\label{fig:delta_sts}}}
\end{figure}

While the RKC scheme presents no oscillations in the solution for $l=100$ time-steps regardless of the upwinding or $l=10$ time-steps with appropriate upwinding, this is not true of the RKL scheme without damping shift for very low number of time-steps such as $l=10$ (Figure \ref{fig:delta_sts}). The RKG scheme, with its increased stability region fares better on this example. 

In a nutshell, proper upwinding helps significantly to avoid explosions for super-time-stepping schemes, but there are cases where strong oscillations are present in the solution. On this example however, the number of time-steps where oscillations manifest is not really practical. A more careful discretization (less extreme concentration of points near $v=0$) would also help to avoid those oscillations. This is another difference between \citep{foulon2010adi} and \citep{lefloch2019numerical}.

\section{Instability on Black-Scholes}
It is possible to observe a similar phenomenon on the simpler use case of the Black-Scholes model, using a small volatility $\sigma=2\%$ and large interest  rate $r=10\%$. Those settings are somewhat unconventional, as a high return usually goes together with a more moderate volatility. We further consider an expiry barrier option which pays \$1 if the asset spot price $X(T)$ at maturity $T=1$ year is between 10 and 100 and zero otherwise, assuming $X(0)=100$. This is a manufactured contract to make the problem more visible, but the issue may arise with simple vanilla options, although then even more extreme Black-Scholes parameters must be used. We discretize the PDE with central differencing, except at the boundaries $x_{\min}=0$ and $x_{\max}=150$ on a non-uniform grid as in \citep{lefloch2014tr}.

\begin{figure}[h]
	\centering{
		\subfloat[RKL.]{
			\includegraphics[width=.33\textwidth]{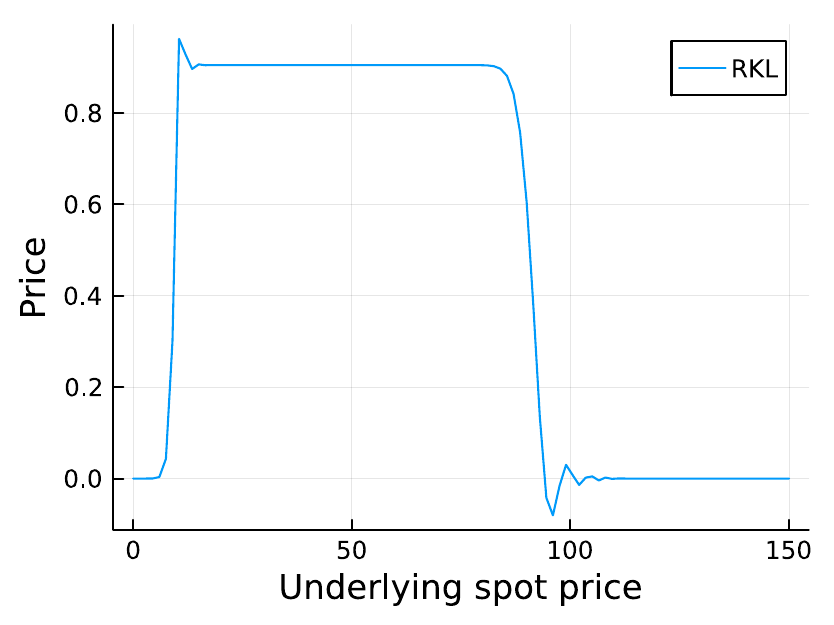}}
		\subfloat[RKG or TR-BDF2.]{
			\includegraphics[width=.33\textwidth]{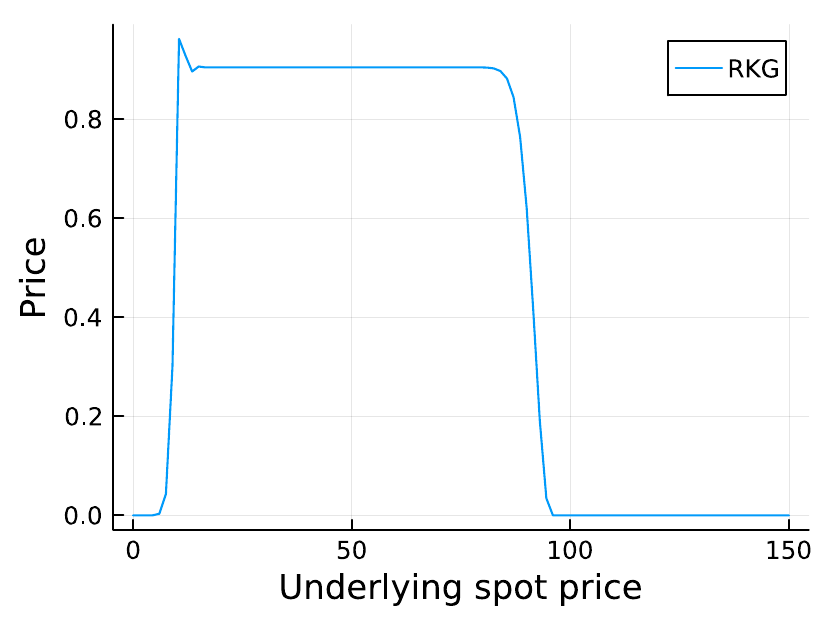}}
		\subfloat[Eigenvalues.]{
			\includegraphics[width=.33\textwidth]{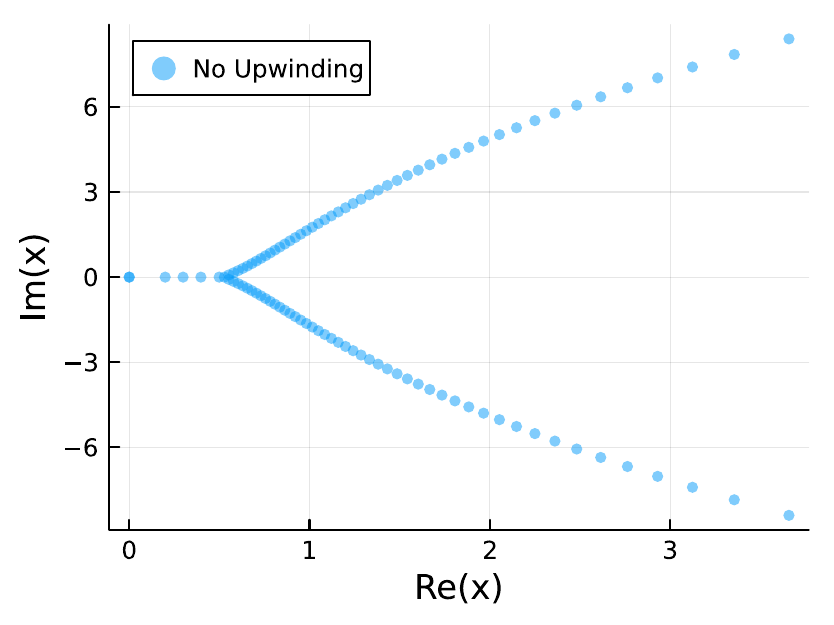}}		
		\caption{Price of an expiry barrier obtained by different super-time-stepping schemes for $l=100$ time-steps  on a uniform grid with $m=100$ space steps. Spurious oscillation around $K$ appear with the RKL scheme. No upwinding is used.\label{fig:delta_bs_uniform_sts}}}
\end{figure}

\begin{figure}[h]
	\centering{
		\subfloat[$P$.]{
			\includegraphics[width=.33\textwidth]{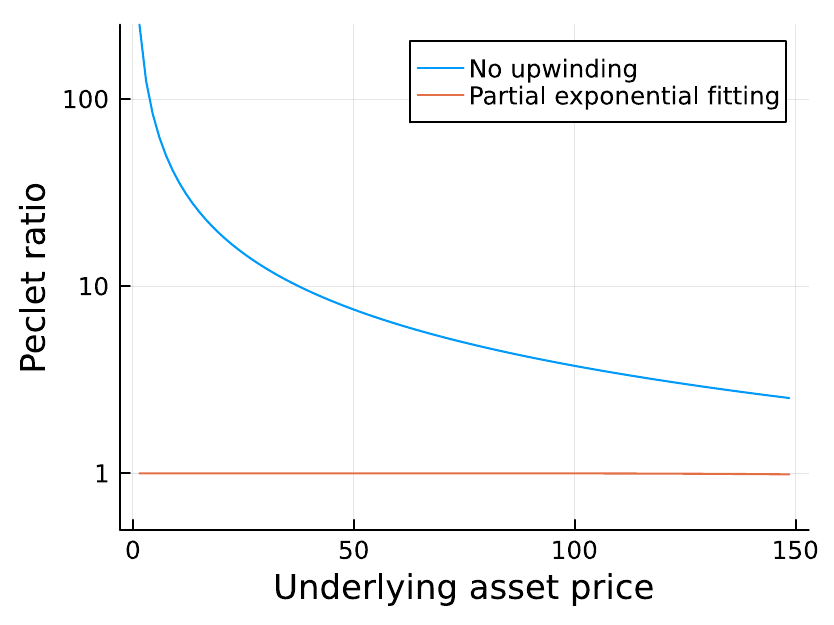}}
		\subfloat[RKL.]{
			\includegraphics[width=.33\textwidth]{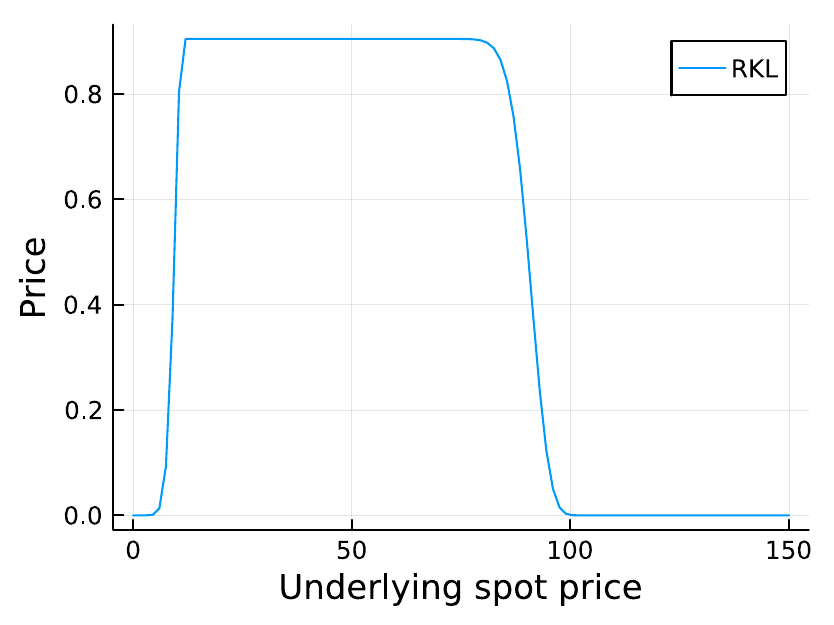}}
		\subfloat[Eigenvalues.]{
			\includegraphics[width=.33\textwidth]{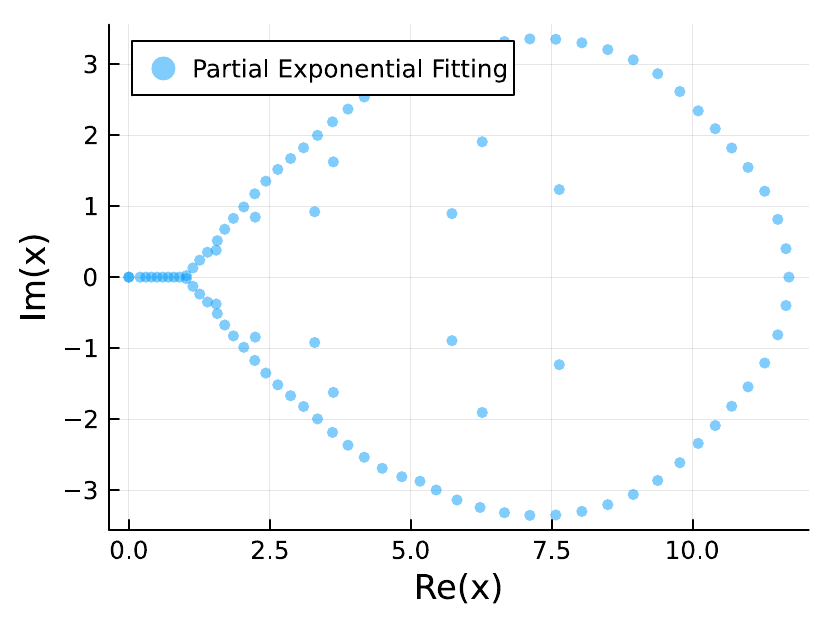}}		
		\caption{Price of an expiry barrier obtained by different super-time-stepping schemes for $l=100$ and $m=100$. Partial exponential fitting is used\label{fig:delta_bs_uniform_exp_sts}}}
\end{figure}

 Figures \ref{fig:delta_bs_uniform_sts} and \ref{fig:delta_bs_uniform_exp_sts} present the option price using a uniform grid respectively without upwinding and with partial exponential fitting.
The option price computed with RKL scheme presents additional oscillations, not present with the TR-BDF2 or RKG schemes, without exponential fitting. Exponential fitting removes any oscillation.

\begin{figure}[h]
	\centering{
		\subfloat[RKL.]{
			\includegraphics[width=.33\textwidth]{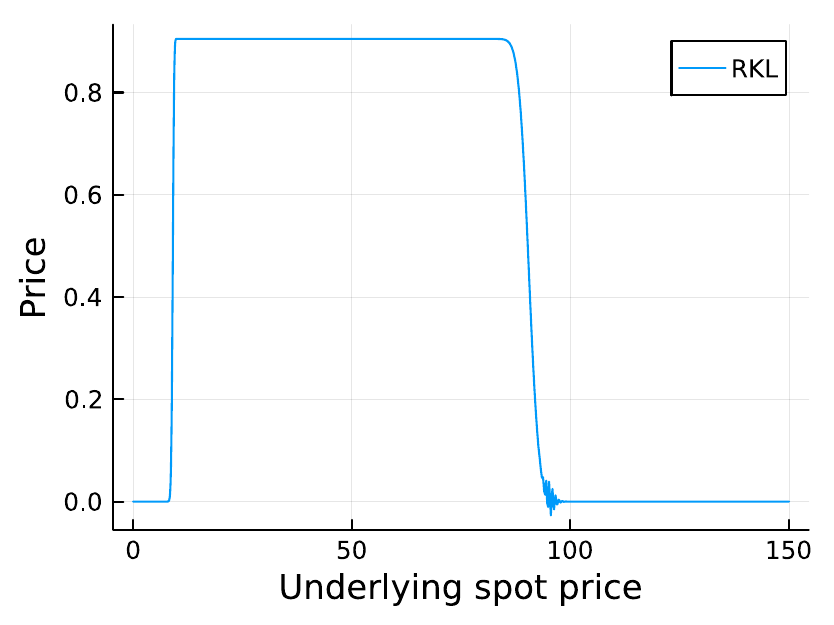}}
		\subfloat[RKG or TR-BDF2.]{
			\includegraphics[width=.33\textwidth]{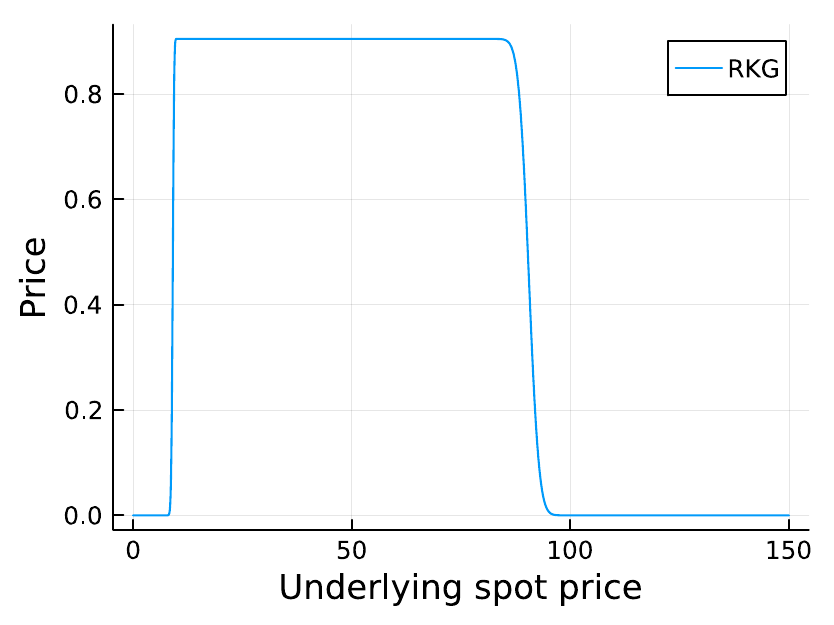}}
		\subfloat[Eigenvalues.]{
			\includegraphics[width=.33\textwidth]{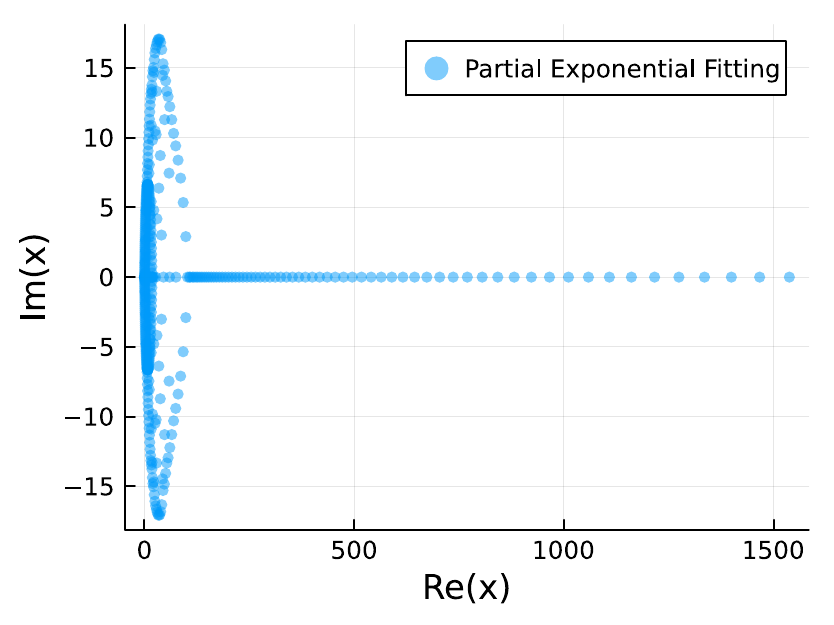}}		
		\caption{Price of an expiry barrier obtained by different super-time-stepping schemes for $l=20$ time-steps  on a cubic stretched grid with $m=400$ space steps. Spurious oscillation around $K$ appear with the RKL scheme. Partial exponential fitting is used.\label{fig:delta_bs_streched_sts}}}
\end{figure}

Figure \ref{fig:delta_bs_streched_sts} shows the price on a cubic stretched grid with stetching coefficient $\alpha=0.01$ \citep{healy2022inserting}.
Small oscillations are visible with the RKL scheme and a low number of time-steps such as $l=20$. They disappear if we increase the number of time-steps (for example to $l=50$) and are absent with the RKG and TR-BDF2 schemes. 

On this simple one-dimensional problem, there is however no explosion of the option price. The problem is mostly oscillations which degrade the accuracy of the scheme.

\section{Conclusion}
We have shown that the RKC scheme with large shift works with no oscillations on the more challenging case 2 of \citet{foulon2010adi} where diffusion is small in one of the dimensions and advection large, when upwinding is carefully applied. The RKL scheme presents strong oscillations near $v=0$ for low number of time-steps (below ten), while the RKG scheme presents very mild oscillations under those conditions. 

In general RKL and RKC super-time-stepping schemes may become unstable when advection dominates, more so than explicit Euler. Upwinding is then particularly important and helps significantly, but there are still cases where super-time-stepping schemes may create spurious oscillations or even explode for low number of time-steps.

Non-uniform grids which concentrate points around specific regions are more problematic for super-time-stepping methods: they may create eigenvalues with relatively large imaginary parts, increasing the likelihood of a blow-up with small or moderate numbers of time-steps, and from a performance point of view, many more stages will be needed for stability.

\acknowledgments{The author would like to thank Karel In't Hout. His precise feedback on previous publications motived this note.}

\externalbibliography{yes}
\bibliography{unstable_sts.bib}
\appendixtitles{no}

\end{document}